# Access Control for Hierarchical Joint-Tenancy

JONATHAN KEIRRE ADAMS
Nova Southeastern University
Ft. Lauderdale, Florida
United States
jonaadam@nova.edu

BASHEER N. BRISTOW
IBM Tucson
Tucson, Arizona
United States
bnbristo@us.ibm.com

*Abstract:* - Basic role based access control [RBAC] provides a mechanism for segregating access privileges based upon users' hierarchical roles within an organization. This model doesn't scale well when there is tight integration of multiple hierarchies. In a case where there is joint-tenancy and a requirement for different levels of disclosure based upon a user's hierarchy, or in our case, organization or company, basic RBAC requires these hierarchies to be effectively merged. Specific roles that effectively represent both the users' organizations and roles must be translated to fit within the merged hierarchy to be used to control access. Essentially, users from multiple organizations are served from a single role base with roles designed to constrain their access as needed.

Our work proposes, through parameterized roles and privileges, a means for accurately representing both users' roles within their respective hierarchies for providing access to controlled objects. Using this method will reduce the amount of complexity required in terms of the number of roles and privileges. The resulting set of roles, privileges, and objects will make modeling and visualizing the access role hierarchy significantly simplified. This paper will give some background on role based access control, parameterized roles and privileges, and then focus on how RBAC with parameterized roles and privileges can be leveraged as an access control solution for the problems presented by joint tenancy.

## 1 Introduction

The work presented in this paper presents a solution to the problem of governing access to data in systems that are shared by multiple independent organizations. Often, organizations may have to share access to a particular computer system, yet be required to protect access to certain data contained in that system from certain modes of access by certain types of users in the respective groups that have access to it. We propose a method to utilize Role Based Access Control to implement an access control system for such a scenario. The scenario that this paper deals with has a few unusual requirements. First, due to the fact the systems to be protected are shared access systems to multiple organizations, and RBAC is an hierarchical access control system, it is important to make sure that any access control system implemented is able to take into account the different roles that will exist in the individual organizations' hierarchies. To accomplish this, we propose utilizing a parameterized approach [1] to RBAC.

First we will present the problem. We will examine potential solutions based upon the existing literature. Next we will describe our approach to a solution and possible future work in the area to further refine our results.

## 2. The Problem

Key to the problem is that on one level, the access control hierarchy should represent the roles within the given organization(s). However, in this configuration, there will be multiple hierarchies, representing multiple organizations, each with its own set of users and roles. The overall set of permissions will be the same as the objects to be manipulated will be the same for all of the groups in this situation.

### 2.1. Joint Tenancy

From the perspective of computer systems, joint tenancy is collocation of multiple organizational entities in which data and system resources are shared. This situation can occur for a variety of reasons, for example a scenario where two organizations are clients of a third organization. Both clients may have access to shared computer resources that are owned by the third. Data may be owned by all organizations involved. In such a case, the onus is on the organization responsible for the shared computing resources to take necessary steps to protect the clients' data from misuse and to protect the clients' ability to securely utilize the provided computing resources, which could be compute time on a server [12], storage resources, documents [13], or other shared resources, in the parlance of access control, regarded as objects.

Within the scope of each organization, or tenant, it is possible to consider a hierarchical access control structure. It is possible that individuals within each organization with different access needs based upon on their organizational roles may be required to access the resources in which joint tenancy is occurring. As such, role based access control seems to be an obvious potential solution to such an access problem. RBAC is a proven access control methodology for defining access control policies in a hierarchical context.

## 2.2. Access Hierarchies in a Joint Tenancy Environment

To solve the problem of providing an efficient and effective access control solution in a joint tenancy environment, it is beneficial to combine the benefits of role based access control with a means for managing roles and privileges across organizations. This work seeks to extend prior work involving parameterized roles and privileges to solve the problem of merging access hierarchies, particularly in the context of joint-tenancy.

## 2.3. Role Based Access Control

Role-Based Access Control is a flexible access control model, primarily notable for its addition of the notion of roles into access control. Ferriaolo and Kuhn have said that RBAC is a Mandatory Access Control [MAC] model, but that it diverges from MAC in several ways [9]. Many models have been derived from RBAC to solve a variety of access control related issues. Role Based Access Control has evolved significantly over the past decade, gaining acceptance. The notion of roles in RBAC provides significant benefit to managing access control in enterprise environments. RBAC allows the management of privileges by roles instead of subjects and supports the principle of least privilege and separation of duties [4][10]. The primary benefit of RBAC is that it simplifies user privilege assignment through its use of roles for determining privilege assignments [11].

A significant benefit to the underlying paradigm of RBAC is its adaptability to various computing configurations and constraints used to determine access control decisions. Researchers have adapted RBAC for geo-spatial applications, for ubiquitous computing applications, as well as extending aspects of the model to support features such as cascading delegation, negative permissions, and a variety of others that make the model more flexible and/or easier to manage in an actual implementation. As the uses for computing and data increase in diversity, access control models have been adapted, significantly increasing in complexity. This is particularly true in pervasive computing and distributed computing applications. Because of the nature of these computing applications, many assumptions used in crafting an access control model must be reconsidered. A significant part of the added complexity comes in the form of considering other factors. In the case of this work, the goal is to find a suitable extension of RBAC that is adaptable to the constraints involved in web services, while adding the benefits derived from a parameter based approach to specifying roles.

## 2.4. Parameterized Roles and Privileges

In [3], parameterization is adding descriptive information to roles and privileges to allow greater flexibility. In terms of privileges, [3] describes parameterized privileges as 2-tuples, where the first parameter represents the object to be accessed and the second parameter represents the access mode for that object given the privilege. With an object *customerData*, and permissions read, write, and null, example privileges would look like:

·$Priv_1$ = *(customerData, read)*
·$Priv_2$ = *(customerData, write)*
·$Priv_3$ = *(customerData, null)*

Thus, privileges are of the form $Priv_n = (o, p_n)$, where p is a set of privileges for an object *o*. Parameterization also affects the subject portion of the RBAC equation. In [3], the authors describe roles with the notation (*rname, rpset, rparamset*) where *rparamset* is the set of role parameters for a given role with the name *rname*. The *rparamset* can be a null value if the role has no given parameters. An example role *customerDataService* that has an identifying attribute *className* with possible values *updateCustData*, *readCustData* and, *eraseCustData* could be represented as:

· *(customerDataService, className, updateCustData)*
· *(customerDataService, className, readCustData)*

· *(customerDataService, className, eraseCustData)*

In [3], it is notable that *PA*, the privilege to role assignment, is represented using XML XPath notation. Because of this, a privilege to role assignment example incorporating the above-explained notation could be represented as:

· $pa_1$ = *(//customerDataService[@className = updateCustData]/customerDataService,update)*
· $pa_2$ = *(//customerDataService[@className = updateCustData]/customerDataService,read)*
· $pa_3$ = *(//customerDataService[@className = updateCustData]/customerDataService,erase)*

For this work, since it will be using XACML, the approach will differ. XACML links *PolicySets* using the <*PolicySetIdReference*> tag. In XACML 1.x, RBAC is implemented using linked *PolicySets* in this manner. Allowing for parameterization will simply require more complicated links between *PolicySets*.

### 2.4.1 Support for Role Based Access Control in XACML

It is notable that as of the publication time of SRBAC, the XACML standard did not directly support the notion of roles, [5] which are essential in RBAC and RBAC derived access control models. In the meantime, support has been added for roles in the form of an XACML profile for role-based access control in 2004. This document is specifically designed to extend XACML such that it will fulfill the requirements necessary to support RBAC without changes to the 1.0 or 1.1 standard [7].

With the second release of the XACML specification, a new profile for role based access control has been specified. This profile includes the necessary support for roles, but specifically supports core RBAC, also called $RBAC_0$ and hierarchical RBAC, also called $RBAC_1$ [8]. Within the context of this work, adding parameters to roles does not seem to violate any aspect of $RBAC_0$ or $RBAC_1$, although in the implementation, there may be some issues involved in formatting parameterized roles such that they will fit into the XACML implementation framework of our system.

For the implementation phase of this project, the author chose Sun Microsystems' XACML implementation. This is the only widely available open source implementation of XACML. Sun's implementation is Java-based, which helps to make it the ideal choice for implementing a web-service or web service enabling library. Sun's XACML code also requires an external XML parser, due to some limitations in the parser (Crimson) that it is included in Java 1.4.x. Among other things, the Crimson parser does not support XPath.

### 2.4.2 Utilizing Parameterized Roles for RBAC in XACML

Working with XACML version 1, the notion of Roles in RBAC are best handled in way that this paper will describe as policy layering. One policy is used to determine which subjects belong to a role. Other policies can be linked, and through them, privileges can be derived. An example of this can be found in the Appendix. There are examples of specifying roles, assigning privileges to roles, as well as sample requests and their responses given the configuration described in the XACML markup in the appendix.

## 2.5 Management of Policies and Roles

In any access control system, the ability to manage policy and, in the case of role based access control, subject-role assignment constitutes critical functionality that directly contributes to the functionality and usefulness an access control system. One of the primary benefits of role-based access control is the ability to manage a large set of policies and users due to the ability to positively or negatively assign permissions based upon a user's role. In this work, we will need to find a way to manage users and roles that have the potential to contain parameters that affect the role or roles assigned and thus the permission set of users. As explained in [2], the ability to have flexible roles and permissions based upon parameters assigned to users can significantly simply managing role permissions, as a single role can, in some situations replace multiple roles.

### 2.5.1 Managing Roles

Because XACML, even in its latest standard version, 2.0, does not support the administration of role based access control, management of roles and policy will not be treated in this work. It is possible to craft a framework for managing roles and policies, but without a standardized framework for implementing it, such as XACML, such a framework would be too unwieldy and time consuming for the purpose of this research. It is likely that managing parameterized roles will require some extension of base role based access control management facilities inherent in any standard that is released by OASIS. Depending upon the next iteration of the XACML standard, this area may be one that is a subject for future work.

# 3. The Solution

We believe that RBAC provides a flexible potential solution to the problem we describe in this paper. With role based access control and parameterized roles and privileges, as described in [2] and [3] and [11] it is possible to represent the merged hierarchical structure that is necessary to control access in a joint-tenancy environment. We view each organization as tree structures of roles and privileges. We use parameterization of roles to merge multiple hierarchies of roles into a singular role-tree. We hope to leverage many of the benefits discussed in [2] of utilizing parameterized role based access control to control access in a service centric environment and modify our approach to suit the not-necessarily service centric, multiple hierarchy, shared object environment of this work.

## 3.1. Merging Access Hierarchies in a Joint Tenancy Environment

In a joint tenancy environment, individual organizations' users and roles can be considered as sets. In our example, we have multiple organizations with unique roles, represented as follows:

*x:* the number of role hierarchies present

*y:* the number of roles in a set

$R_0 = \{ r_{00}, r_{01}, r_{02,...} r_{0y} \}$

$R_1 = \{ r_{10}, r_{11}, r_{12,...} r_{1y} \}$

$R_2 = \{ r_{20}, r_{21}, r_{22,...} r_{2y} \}$

…

$R_x = \{ r_{x0}, r_{x1}, r_{x2,...} r_{xy} \}$

From a hierarchical perspective, we represent an individual set of roles, assuming a hierarchy with four roles, the super role being the *System Administrator* with three subordinate roles with as described in *figure 1*.

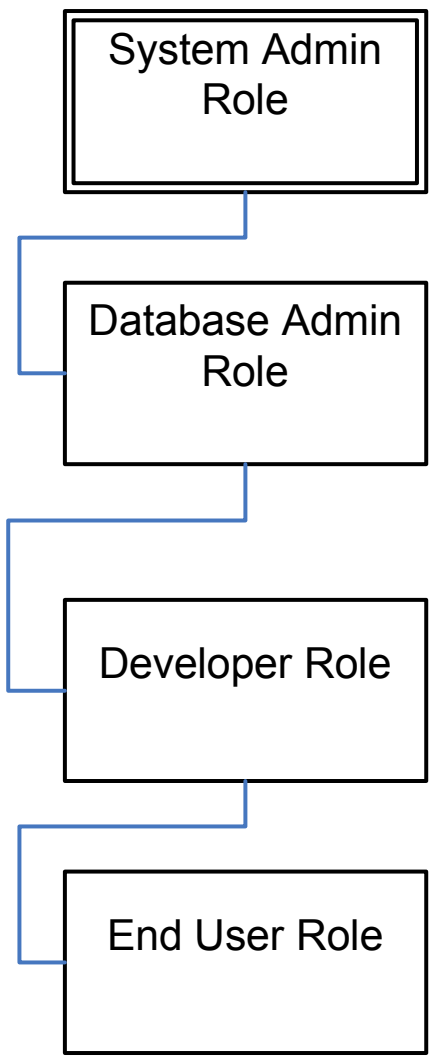

Fig. 1 *Set of Hierarchical Roles*

## 3.2. Implementing RBAC Across a Merged Access Hierarchy

The primary goal of the work described in this paper is to merge the sets of hierarchical roles such where:

*n:* the number of tenants

$R_{0...}\ R_n$: the hierarchical role sets for each tenant,

$R_{all} = R_0 \cup R_1 \cup R_2 \ldots \cup R_n$

In order to accomplish this goal, while maintaining the integrity of the role hierarchies that comprise $R_{all}$, we use parameterized roles.

We are able to pass a unique hierarchy-related value as a parameter to each role. We place all of the roles into the set $R_{all}$. The hierarchy-related value can be used to reconstitute the individual hierarchies as well as represent each role's position for privilege assignment. Being able to maintain the merged list of roles and keeping the hierarchical information, it is possible to generate parameterized permissions that are able to take into account the source and level of the respective role or roles as access decisions are made. In our example, the parameterized roles can be represented as a three-tuple of the form:

*(roleName,rparamset,,hierarchyIdentifier)*

*roleName* is the identifier for the role, which should be unique to it within its individual hierarchy and *hierarchyIdentifier* is a unique identifier for a specific hierarchy. *Rparamset* is a set of parameters, but in our case, the set contains a single value to signify that the set of parameters should contain the *hierarchyIdentifier*.

The ability to utilize parameterized roles to solve joint-tenancy related access control problems is bolstered by the existence of the XACML standard. XACML provides direct support for role based access control and indirect support for parameterized roles and privileges as shown in [1]. XACML gives us a means to represent policies, as well as access decisions using XML markup.

The decision to about whether to use parameterized permissions is based upon a couple of factors. In this work we do not spend significant coverage on the topic.

# 4. Conclusion

In this work, we have presented the problem of managing hierarchical roles and privileges in an environment of joint tenancy. We propose a solution that utilizes the benefits of Role Based Access Control with parameterized roles and privileges. We formally frame the problem and give examples how such a solution may be formulated from a conceptual perspective.

As potential further work, we propose work more work to be done toward a formal RBAC based access control model to target the problem of access control in a joint-tenancy environment. We propose that a mock up system be developed, particularly leveraging standards such as XACML and utilizing Sun Microsystems XACML framework, as it provides excellent support for RBAC and has been proven capable of representing parameterized roles and privileges [1].

# 5. Appendix

## 5.1. Sample XACML Role to Privilege Assignment (Role Definition)

```
<PolicySet xmlns= "urn:oasis:names:tc:xacml:1.0:policy"
PolicySetId="RPS:physician:role"
PolicyCombiningAlgId= "urn:oasis:names:tc:xacml:1.0:policy-combining-algorithm:permit-overrides">
<Target>
<Subjects>
<Subject>
<SubjectMatch MatchId=
"urn:oasis:names:tc:xacml:1.0:function:anyURI-equal">
<AttributeValue DataType=
"http://www.w3.org/2001/XMLSchema#anyURI">urn:example:rolevalues:
physician</AttributeValue>
<SubjectAttributeDesignator
AttributeId="urn:oasis:names:tc:xacml:1.0:subject:role"
DataType="http://www.w3.org/2001/XMLSchema#anyURI"/>
</SubjectMatch>
</Subject>
</Subjects>
</Target>
<PolicySetIdReference>PPS:physician:role</PolicySetIdReference>
</PolicySet>
```

## 5.2. Sample XACML Permission Assignment (Policy)

```
<?xml version="1.0" encoding="UTF-8"?>
<PolicySet xmlns="urn:oasis:names:tc:xacml:1.0:policy"
PolicySetId="PPS:physician:role"
PolicyCombiningAlgId="urn:oasis:names:tc:xacml:1.0:policy-combining-algorithm:permit-overrides">
<Target/>
<Policy PolicyId="Permissions:specifically:for:the:physician"
RuleCombiningAlgId="urn:oasis:names:tc:xacml:1.0:rule-combining-algorithm:permit-overrides">
<Target/>
<Rule RuleId="Permission:to:create:prescriptions"
Effect="Permit">
<Target>
<Resources>
<Resource>
<ResourceMatch
MatchId="urn:oasis:names:tc:xacml:1.0:function:string-equal">
<AttributeValue
DataType="http://www.w3.org/2001/XMLSchema#string">prescription</AttributeValue>
<ResourceAttributeDesignator
AttributeId="urn:oasis:names:tc:xacml:1.0:resource:resource-id"
DataType="http://www.w3.org/2001/XMLSchema#string"/>
</ResourceMatch>
</Resource>
</Resources>
<Actions>
<Action>
<ActionMatch MatchId="urn:oasis:names:tc:xacml:1.0:function:string-equal">
<AttributeValue
DataType="http://www.w3.org/2001/XMLSchema#string">create</AttributeValue>
<ActionAttributeDesignator
AttributeId="urn:oasis:names:tc:xacml:1.0:action:action-id"
DataType="http://www.w3.org/2001/XMLSchema#string"/>
</ActionMatch>
</Action>
</Actions>
</Target>
</Rule>
<Rule RuleId="FinalRule" Effect="Deny"/>
<Obligations>
<Obligation
ObligationId="urn:obligation-physician"
FulfillOn="Permit">
<AttributeAssignment
AttributeId="urn:explanation"
DataType="http://www.w3.org/2001/XMLSchema#string">only phyisicans can create prescriptions
</AttributeAssignment>
</Obligation>
</Obligations>
</Policy>
</PolicySet>
```